\documentclass[]{interact}

\usepackage{epstopdf}
\usepackage[caption=false]{subfig}
\usepackage{color}
\usepackage[
colorlinks,
linkcolor=blue,
urlcolor=blue,
citecolor=blue
]{hyperref}
\usepackage{siunitx}

\usepackage[numbers,sort&compress,merge]{natbib}
\bibpunct[, ]{[}{]}{,}{n}{,}{,}

\theoremstyle{plain}

\theoremstyle{definition}

\theoremstyle{remark}

\usepackage{comment}
\usepackage{braket}
\usepackage{booktabs}

\begin{document}

\articletype{Special Issue in Honour of Frédéric Merkt}

\title{Theoretical Calculation of Electron Transfer Between Calcium Ground-State Atoms and Rydberg Atoms}

\author{
\name{A.\,Bouillon\textsuperscript{a}\thanks{CONTACT M. Génévriez. Email: matthieu.genevriez@uclouvain.be} and M.\,Génévriez\textsuperscript{a}}
\affil{\textsuperscript{a} Institute of Condensed Matter and Nanosciences, Universit\'e catholique de Louvain, BE-1348 Louvain-la-Neuve, Belgium
}}

\maketitle

\begin{abstract}
We calculated the electronic interaction associated with the exchange of an electron
between an atom of calcium excited to a Rydberg state ($n\sim 10-15$) and another,
neighbouring calcium atom in its ground state. In this range the Rydberg states
have an energy that is comparable to the electron affinity of Ca, enabling resonant or
near resonant charge transfer at large internuclear separations (200-700 $a_0$). We calculated the interaction strength while systematically and critically assessing the approximations made, and found it to be large, ranging from $10^{-5}$ \textcolor{black}{$E_h$} (70 GHz) to $10^{-8}$ \textcolor{black}{$E_h$}. Charge transfer is thus expected to be efficient and to significantly affect the molecular dynamics at a range of internuclear distances where ultralong range Rydberg molecules also exist.
\end{abstract}


\section{Introduction}

Electron transfer (ET) is a fundamental process of chemistry that takes place,
in the gas phase, when two atomic or molecular compounds A and B
scatter at distances close enough to exchange an electron, thereby yielding a
positive ion (A$^+$) and a negative ion (B$^{-}$). Owing to its
fundamental nature, electron transfer has been the subject of extensive
experimental studies and numerous theoretical works \cite{mollet10,beyer18,peper20,hummel20} . The reverse process of mutual neutralization, whereby an electron is transferred from B$^-$ to A$^+$ upon collision, has also been extensively studied and a great level of detail is being reached thanks to the recent development of cryogenic storage rings (see \cite{bogot24,poline24,poline25,mull21} and references therein).

\begin{figure}
	\includegraphics[width=\columnwidth]{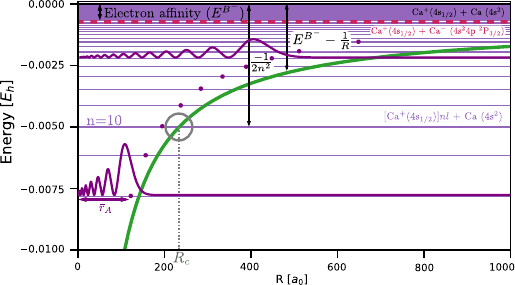}
	\caption{Potential energy curves of interest in the case where both A and B are Ca atoms. The crossing internuclear distance $R_c$ is highlighted by a gray empty circle for the $nl=10f$ Rydberg state. \label{fig:Energies} }
\end{figure}

The short distance at which electron transfer takes place in most atomic and
molecular systems is the result of two constraints. First, the electronic
wavefunction of A must significantly overlap the one of B$^-$ because the ET
strength is related to the matrix element $\braket{\psi_A | H | \psi_{B^-}}$,
with $H$ the total electronic Hamiltonian. Second, the energies of one electronic state of the pair (A, B) must be close to the one of (A$^+$, B$^-$) such that the transfer is resonant or near resonant. It implies that, considering only the strongest, Coulomb interaction between A$^+$ and B$^-$,
\begin{equation}
	E^{B^-} -\frac{\textcolor{black}{e^2}}{4\pi\epsilon_0 R} \simeq  E^A .
\end{equation}
If the negative ion is in its ground state, $E^{B^-}$ is minus the electron affinity of B, $R$ is the internuclear distance, and $E^A$ the binding energy of the electronic state of A under consideration. The above equation imposes in turn a constraint on the typical sizes $\bar{r}_A$ and $\bar{r}_{B^-}$ of the electronic wavefunctions of A and B$^-$, respectively,
\begin{equation}
	\bar{r}_A + \bar{r}_{B^-} ~\textcolor{black}{\simeq}~ R_c = \frac{\textcolor{black}{e^2}}{4\pi\epsilon_0(E^{B^-}-E^A)} .
\end{equation}
In most cases (i) the electron affinity of B (\textcolor{black}{$E^{B^-}$}) is vastly
different from $E^A$ and $R_c$ is small, typically of the order of a few Bohr
radii, or (ii) even if $E^{B^-} \sim E^A$ and $R_c$ is large, $\bar{r}_A$ and
$\bar{r}_{B^-}$ are both small such that electron transfer takes place
nonetheless at short distances. Calcium stands out from these cases because its
electron affinity is very low, in fact one of the lowest of the periodic table (19.73~meV for the $4s^24p~^2P_{3/2}$ term \cite{petrunin96}). The energy range where $R_c$ is large corresponds to electronic
states of A with low binding energies ($E^{B^-} \sim E^A$), i.e., to highly
excited Rydberg states. This situation is illustrated in Fig. \ref{fig:Energies}, where the horizontal full purple lines show the energies $E^A$ of some Rydberg states of Ca whereas the thick full green line shows $E^{B^-} - \textcolor{black}{e^2}/4\pi\epsilon_0 R$. The internuclear distance at which they cross is $R_c$. $\bar{r}_A$, which we take as the classical outer turning point of the Rydberg electron trajectory, is also indicated by the full circles. Both $R_c$ and $\bar{r}_A$ are large, which means that electron transfer can take place at distances exceeding
100~$a_0$. More specifically, we estimate that it becomes significant for
Rydberg states with principal quantum numbers $n$ lower than $\sim 16$, which
corresponds to a distance $R_c \simeq \bar{r}_A \simeq 700~a_0$.

Charge transfer between Rydberg atoms and ground-state calcium atoms has been
investigated in the past using thermal beams~\cite{mclaughlin94,reicherts97}.
In such cases experimental and theoretical effort focused on the calculation of
total cross sections and final-state
distributions~\cite{fabrikant00,lorensen96}, convoluted by the thermal
distribution of kinetic energies. Electron transfer was measured to peak around
$n\sim 10$, with good agreement between theory and experiment. Because
theoretical work focused on the collisional aspects of electron transfer, using
for example close-coupling-type calculations, little is known on the structure
of potential energy curves near the charge exchange region. Moreover, in the
same region the potential-energy curves of the (A, B) pair are complex and show
local \textcolor{black}{potential} wells that support vibrational bound states, the so-called
ultra-long-range Rydberg molecules (ULRRMs) \cite{eiles15}. It is the purpose of the present
article to establish the theoretical basis for accurately calculating potential-energy curves and molecular dynamics in this range, with a particular emphasis on accurately computing the matrix
elements governing electron transfer.

Our work builds upon previous theoretical proposals for preparing ion pairs in ultracold alkali-atom gases through photoexcitation of ULRRM. One proposal is to take advantage of the attractive singlet $p$-wave electron scattering in Rb to excite ULRRMs bound in a stairwell potential as a transitional step towards ion pair excitation \cite{hummel20}. Other proposals in Rb are based on ion pairs formation by electric field control of ultracold Rydberg molecules \cite{markson16} or by Feshbach resonances \cite{kirrander13}. A two-step approach was also studied in Cs \cite{peper20}, where the transition between the two molecular systems relies on stimulated emission.

The paper is organized as follows. In Sec. \ref{sec:theory}, the theory relative to the description of the system is presented. The first three subsections are dedicated to the Hamiltonian of the system and the determination of the electronic wavefunctions of the ULRRM and of the negative ion. The calculation of the interaction leading to the charge transfer is then thoroughly described with emphasis put on critically assessing each assumption made along the derivation. In Sec. \ref{Sec:results}, results are discussed considering, for the sake of example, the case of A being a Ca atom in a Rydberg state and B being a ground-state Ca atom. The theory outlined in the first section is however general and calculations can be easily extended to other atomic species for A and other weakly-bound negative ions B$^-$ that can be well described with a single-active-electron framework. Atomic units are used throughout the paper unless stated otherwise.

\section{Theory}
\label{sec:theory}
\subsection{Outline of the problem}\label{sec:problem_outline}
\begin{figure}
	\includegraphics[width=\columnwidth]{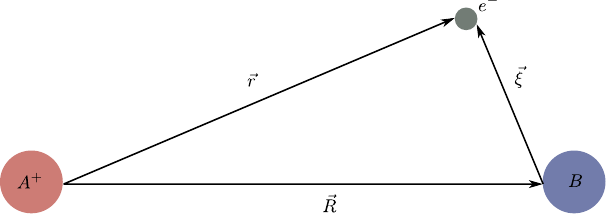}
	\caption{Schematic representation of the problem and the associated coordinates.\label{fig:coord}}
\end{figure}

We start from a system composed of one positive ion A$ ^+$, one electron e$^-$ and one ground-state perturber atom B, as shown in Fig. \ref{fig:coord}. \textcolor{black}{This means that we consider only \emph{one} electron evolving in the potentials of both A$^+$ and B. As the internuclear distance $R$ tends to infinity, the system is either in the [A$^+$]$nl$ + B state or the A$^+$ + B$^-$ state. This is no longer true at shorter internuclear distances, where the electron can transfer from the A$^+$ center to the B center and vice versa.} The Born-Oppenheimer Hamiltonian \textcolor{black}{governing the motion of the electron in the molecule-fixed frame} can be written as \textcolor{black}{\citep{giannakeas20,fey20,eiles19,omont77}}
\begin{align}
H(\bm{r};\textcolor{black}{\bm{R}}) = \textcolor{black}{-\frac{\bm{\nabla_r}^2}{2}} + V_{A^+}^{\text{sh}} (\bm{r})- \frac{1}{r} + V_{B} (\textcolor{black}{|\bm{r}-\bm{R}|}) ,
\label{Hamiltonian}
\end{align}
where \textcolor{black}{$\bm{R}$ is the internuclear distance and enters the Born-Oppenheimer electronic Hamiltonian as a parameter}, $V_{A^+}^{\text{sh}}$ is the short-range potential describing the interaction of the electron with the other electrons and nucleus of A$^+$, $-1/r$ refers to the long-range Coulomb potential due to A$^+$ and $V_{B}$ describes the interaction potential between the electron and the ground-state atom.
For $|\bm{r}|$ sufficiently large and neglecting $V_B$ altogether, the time-independent Schrödinger equation is the one of a Rydberg electron \citep{aymar96},
\begin{align}
\left[\textcolor{black}{-\frac{\bm{\nabla_r}^2}{2}} + V_{{A^+}}^{\text{sh}} (\bm{r}) - 1/r \right] \Phi_{n_1l_1m_{l_1}}^{A}(\bm{r}) = E^{A}_{n_1l_1} \Phi_{n_1l_1m_{l_1}}^{A}(\bm{r}),
\label{eq:schrodinger_A}
\end{align}
with
\begin{align}
\Phi_{n_1l_1m_{l_1}}^{A}(\bm{r}) = \frac{\textcolor{black}{W(r,n^\star_1,l_1)}}{r}~Y_{l_1m_{l_1}}(\theta,\phi).
\end{align}
\textcolor{black}{$W(r,n^\star_1,l_1)$ is the normalized Whittaker Coulomb function of orbital angular-momentum quantum number $l_1$ and effective quantum number $n^\star_1 = n_1-\mu_{n_1l_1}$, where $n_1$ is the principal quantum number and $\mu_{n_1l_1}$ the quantum defect \citep{aymar96}}. $Y_{l_1m_{l_1}}(\theta,\phi)$ is a spherical harmonic of magnetic quantum number $m_{l_1}$.  \\
Conversely, neglecting the potential of A$^+$ gives the time-independent Schrödinger equation for the anion,
\begin{align}
\left[\textcolor{black}{-\frac{\bm{\nabla_\xi}^2}{2}} + V_{B} (\bm{\xi})\right]\psi_{l_2m_{l_2}\alpha_2}^{B^-} (\bm{\xi}) = E^{B^-} \psi_{l_2m_{l_2}\alpha_2}^{B^-} (\bm{\xi}),
\label{anion_wavefunction}
\end{align}
where $l_2$ is the orbital angular-momentum quantum number associated to the orbital motion of the electron around the neutral atom, and $m_{l_2}$ the associated magnetic quantum number. $\alpha_2$ designates any other quantum number necessary to fully specify the state. \\
We describe the electronic wavefunction as a linear combination of functions corresponding to Rydberg states of the Ca atom \textcolor{black}{$\ket{\Phi_{nlm}^{A}}$} and a function corresponding to the anion bound state \textcolor{black}{$\ket{\psi^{B^-}_m}$}. The electronic wavefunction is therefore written, as an \textit{ansatz}, as \textcolor{black}{\citep{markson16,dochainarnaud,poline22}}
\begin{align}
\ket{\Psi_m} = a \ket{\psi^{A}_m} + b \ket{\psi^{B^-}_m}.
\label{Psi_m}
\end{align}
The wavefunction centered on $A^+$ is written, in generality,
\begin{align}
\ket{\psi^{A}_m} = \sum_{nl} c_{nl}\ket{\Phi_{nlm}^{A}},
\end{align}
\textcolor{black}{in order to describe not only isolated Rydberg states $\ket{\Phi_{nlm}^{A}}$ but also the electronic states of ultra-long-range Rydberg molecules, as shall be developed in Sec. \ref{subsec:Rydberg_wavefunction}. We also note that, at this point, the $\ket{\psi^{A}_m}$ and $\ket{\psi^{B^-}_m}$ wavefunctions are expressed with respect to two different centers, a difficulty that will be treated in Sec. \ref{Interaction_term}.}
The projection of the electronic orbital angular momentum along the internuclear axis being a conserved quantity, the magnetic quantum numbers are identical ($m=m_{l_1}=m_{l_2}$).
$a$ and $b$ are complex numbers fulfilling $|a|^2+|b|^2 + 2\Re \{ a^\star b  \braket{\psi^{A}_m|\psi^{B^-}_m} \}=1$. Because the basis functions $\ket{\psi^{A}_m}$ and $\ket{\psi^{B^-}_m}$ are not necessarily orthogonal, we must solve the generalized eigenvalue problem. In the simple case where only one ion-pair state and one ultra-long-range molecular state are sufficient to describe the system, we can write the generalized eigenvalue equation as $\textcolor{black}{\tilde{H}}\ket{\Psi_m} = ES\ket{\Psi_m}$, where \textcolor{black}{$\tilde{H}$ is the matrix containing the matrix elements of $H$ in the $\left\{\ket{\psi^{A}_m},\ket{\psi^{B^-}_m}\right\}$ basis} and $S$ is the overlap matrix defined by
\begin{equation}
S = \begin{pmatrix}
1 & s \\
s^\star & 1
\end{pmatrix} \quad \text{with } s = \braket{\psi^{A}_m|\psi^{B^-}_m}.
\end{equation}
To find the solution one must diagonalise the effective Hamiltonian $H_{\text{eff}}=S^{-1}\textcolor{black}{\tilde{H}}$. Neglecting the terms in $|s|^2$, we obtain
\begin{equation}
H_{\text{eff}}\simeq \begin{pmatrix}
\mathcal{E}^{A} - s\bra{\psi^{A}_m} V_{B} \ket{\psi^{B^-}_m}
& \bra{\psi^{A}_m} V_{B} \ket{\psi^{B^-}_m} + \sigma^\star -s \mathcal{E}^{B^-} \\
\bra{\psi^{B^-}_m} V_{B} \ket{\psi^{A}_m} + \sigma -s^\star \mathcal{E}^A & \mathcal{E}^{B^-} -s^\star \bra{\psi^{B^-}_m} V_{B} \ket{\psi^{A}_m}
\end{pmatrix}
\label{H_eff}
\end{equation}
where we defined
\begin{align}
\mathcal{E}^A &= \sum_{nl}|c_{nl}|^2E_{nl} + \bra{\psi^{A}_m} V_{B} \ket{\psi^{A}_m}, \label{Epsilon_ULRRM}\\
\mathcal{E}^{B^-} &= E^{B^-} - \frac{1}{R},  \quad \text{and}\\
\sigma &= \sum_{nl}c_{nl}E_{nl}\braket{\psi^{B^-}_m|\Phi^{A}_{nlm}}.
\end{align}

The term $\mathcal{E}^A$ corresponds to potential-energy function of the ULRRM, consisting of one Rydberg atom and one ground-state atom onto which the Rydberg electron scatters elastically, as will be developed in Sec. \ref{subsec:Rydberg_wavefunction}. \textcolor{black}{$\mathcal{E}^{B^-}$ is the potential-energy function of the ion-pair state. We assumed that the variation} of the Coulomb potential across the range of distances over which $V_{B}$ is significant is negligible. In other words, $\bra{\psi^{B^-}} \left( V_{A^+}^{\text{sh}} - \frac{1}{r} \right) \ket{\psi^{B^-}} \sim -1/R$, with $R$ the internuclear distance. Going beyond this assumption would require including the electric-potential gradient in the anion region $(\propto 1/R^2)$. \\
\textcolor{black}{The calculation of the off-diagonal elements of Eq. (\ref{H_eff}) can be done numerically using the approach detailed in Sec. \ref{Interaction_term} to compute the relevant integrals. We note that, if} (i) $\ket{\psi^{A}_m}\simeq\ket{\Phi^{A}_m}$, (ii) the interaction term $\braket{\psi_m^A | V_B | \psi_m^{B^-}}$ is small, as shall be shown later, and (iii) that $R=R_c$, one can show that the terms $\sigma^\star - s\mathcal{E}^{B^-}$, $\sigma - s\mathcal{E}^{A}$ vanish and the terms $\braket{\psi_m^A | V_B | \psi_m^{B^-}}$ are negligible. We then recover the expression of the Hamiltonian in the limit where $\psi_m^A$ and $\psi_m^{B^-}$ are orthogonal. Away from $R_c$, the overlap should be taken into account as soon as $s\left|\frac{R-R_c}{R_c^2}\right|\sim \bra{\psi^{A}_m} V_{B} \ket{\psi^{B^-}_m}$. Beyond those assumptions, in the general case, the overlap must be accounted for. \\

\subsection{Rydberg wavefunction}
\label{subsec:Rydberg_wavefunction}
The $\bra{\psi^{A}_m} V_{B}(\bm{\xi}) \ket{\psi^{A}_m}$ integral in Eq. (\ref{Epsilon_ULRRM}) corresponds to
the elastic scattering of the Rydberg electron off the calcium ground-state
perturber atom. Multiple scatterings of this electron mediate a binding
mechanism between the ground-state atom and the ion, leading to the formation
of a slightly bound, ultra-long-range Rydberg molecule (ULRRM) \cite{greene00,bendkowsky09}. This can be
described, in zero-range approximation, by the Fermi-Omont contact
pseudopotential \cite{eiles15,omont77}
\begin{equation}
\begin{aligned}
V(\bm{r},\bm{R}) = 2\pi a_s[k(R)] \delta(\bm{r}-R\hat{z}) + 6 \pi a_p[k(R)] \delta(\bm{r}-R\hat{z}) \nabla_{\bm{r}} \cdot \nabla_{\bm{r}}
\end{aligned}
\label{Fermi_omont}
\end{equation}
assuming, without losing generality, that $\bm{R}$ is aligned with the $z$-axis. $a_s[k(R)]=-\tan (\delta_0) / k(R)$ and $a_p[k(R)]=-\tan (\delta_1) / [k(R)]^3$ are, respectively, the energy-dependent scattering length and scattering volume corresponding to the s- and p-wave scatterings at semiclassical momentum $k(R)= \sqrt{\frac{-1}{\textcolor{black}{n^\star}^2} + \frac{2}{R}}$ \citep{eiles15}. In the case of a Ca neutral perturber, the energy-dependent phase shifts $\delta_0$ and $\delta_1$ can be extracted from the theoretical calculations reported in Ref. \cite{Yuan90} and the quantum defects from \cite{kramida99}. The potential energy curves of the ULRRM states can be computed by diagonalizing (\ref{Fermi_omont}) in a basis of Rydberg wavefunctions $\ket{\Phi^A_{n_1 l_1 m_{l_1}}}$ \textcolor{black}{\citep{peper21}}, or using advanced methods described, for instance, in Refs.~\cite{giannakeas20,eiles15}.

\begin{figure}[h]
	\includegraphics[width=\columnwidth]{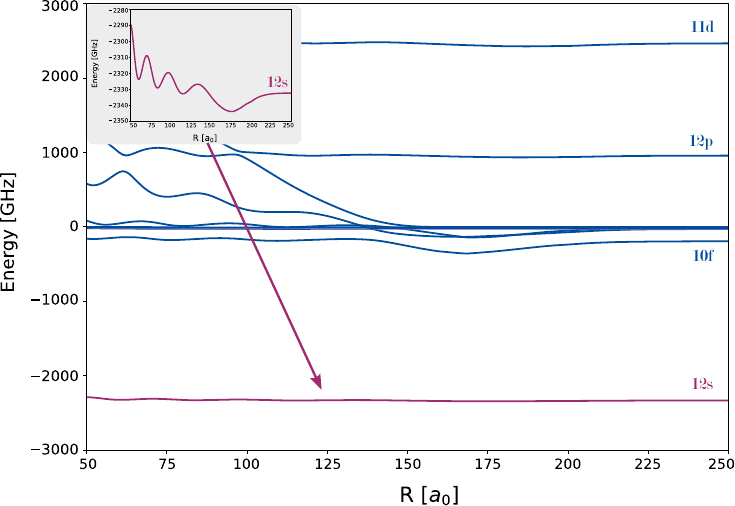}
	\caption{Potential-energy functions of ULRRMs associated to the $n=10$ manifold ($m=0$), with the zero-quantum-defect Rydberg state taken to be the reference zero energy. The small included panel zooms in on one of the ULRRM state, associated to the $12s$ Rydberg level of the isolated atom. \label{fig:ULRRM} }
\end{figure}

Figure \ref{fig:ULRRM} shows the ULRRM potential-energy functions for Ca atoms in the region of effective principal numbers $n^* \sim 10$ computed with the diagonalization method mentioned above. All Rydberg states belonging to the $n-1,n+1$ and $n+2$ manifolds were also included in the diagonalization to improve the convergence of the potential-energy functions. The ULRRM states are designated by the atomic Rydberg state towards which they converge at large internuclear distances. From this diagonalization, we also retrieve the $c_{nl}$ coefficients necessary to build the wavefunction $\ket{\psi^{A}_m}$.

The zoom, in Fig. \ref{fig:ULRRM}, on the ULRRM potential-energy functions converging towards the $12s$ Rydberg state stresses that, at such small $n$ values, the attractive potential arising from the s-wave scattering is substantially deeper than what is usually achieved in experiments involving ULRRM \cite{desalvo15,bendkowsky09,booth15}, with a depth of \textcolor{black}{$\sim 10$} GHz for the outermost well. This is explained by the fact that we are considering lower $n$ values than in typical experiments.

\subsection{Anion wavefunction}
\label{sec:anion_wvfn}
The anion wavefunction $\ket{\psi^{B^-}_m}$ in Eq. (\ref{anion_wavefunction}) is the one of a bound state, and often the only bound state, of the potential $V_B$. At sufficiently large radial distance, the interaction potential $V_B$ between an electron and the neutral atom B is a polarisation potential taking the form
\begin{equation}
V_B(\xi) = \frac{-\alpha}{2\xi^4} \left( 1 - \exp \left( \frac{\xi}{\xi_0} \right)^6 \right) , \label{eq:vb}
\end{equation}
with $\xi = \lVert \bm{\xi} \rVert$ being the radial distance between the electron and the perturber, \textcolor{black}{$\alpha=159.4$ a.u. \citep{mitroy08}} the polarisability of the Ca ground state atom and $\xi_0$ a constant to be determined. To do so, the radial part of equation (\ref{anion_wavefunction}) was numerically solved \textcolor{black}{for $l_2 = 1$} with a DVR method (400 grid points, box size of 180 $a_0$) \cite{rescigno00,genevriez19}, and $\xi_0$ optimized so that the resulting energy of the bound state matches $E^{B^-}$=~19.73~meV \cite{petrunin96}. \textcolor{black}{We consider only the binding energy of the $^2P_{3/2}$ spin-orbit component of the ground state} for the calculations presented below and we neglect the spin-orbit coupling. The $^2P_{1/2}$ component has a higher binding energy \textcolor{black}{\citep{petrunin96}}, yielding a larger \textcolor{black}{value of the internuclear distance $R_c$ at which the crossing takes place,} and therefore a lower interaction with ULRRM states. The value of $\xi_0=1.226522\alpha^{1/4}$ used in all subsequent calculations yields an agreement between the calculated and theoretical energies of 4 significant digits. This method therefore led to the determination of both the potential $V_B$ and the wavefunction $\psi^{B^-}(\bm{\xi}) = \chi^{B^-}(\xi) Y_{l_2m_{l_2}}(\theta_\xi, \phi_\xi)$. The resulting wavefunction is in good agreement with the one calculated with many-body perturbation theory in Ref. \citep{dzuba97}. A comparison between the values of the interaction given by this present anionic wavefunction and the one of Ref. \cite{dzuba97} will be shown at the end of Sec. \ref{Sec:results}.

\subsection{Interaction term}
\label{Interaction_term}
The interaction term
\begin{equation}
	\bra{\Phi_{n_1l_1m}^{A}} V_{B} \ket{\psi^{B^-}_{l_2m}}
	\label{eqref:interaction_term_plain}
\end{equation}
entering Eq. (\ref{H_eff}) is the crucial quantity to model electron transfer. The direct, numerical integration of the above integral to determine the value of the matrix element is usually avoided because it is demanding computationally, in particular if one wishes to evaluate it at many internuclear distances for many states $\ket{\Phi_{n_1l_1m}^{A}}$.

The problem is often circumvented by transforming the volume integral into a surface integral under a few, reasonable assumptions \textcolor{black}{\citep{barklem21,dochainarnaud,robicheaux15,janev72}}. As detailed, for example, in Ref.~\citenum{giannakeas20}, the principle of the transformation consists in rewriting the wavefunction $\Phi_{n_1 l_1 m}^A$ centered on the cation $A^+$ into a series of regular spherical Bessel functions centered on the perturber neutral atom $B$. In a first step, the Whittaker function is re-expressed as a sum of two linearly independent, energy-normalized, regular and irregular spherical Bessel functions [$f_{l_1m}$ and $g_{l_1m}$, respectively], with coefficients calculated such that this is valid in the small region around the perturber B. The Bessel functions are then re-expressed in terms of a center located at the $B$ nucleus. This transformation of coordinates is called by some authors~\cite{giannakeas20,giannakeas16} the local frame transformation (LFT). The function $\Phi_{n_1l_1m}^{A}$ is finally written, in the region near B, as
\begin{equation}
\begin{aligned}
\Phi_{n_1l_1m}^{A}(\bm{r} = \bm{\xi} + \bm{R}) &= \sum_{l_2m_{l_2}}  f_{l_2m_{l_2}} (\bm{\xi}) \mathcal{V}^T_{l_2m_{l_2},l_1m} (R), \quad \text{with}\\
\mathcal{V}^T_{l_2m_{l_2},l_1m} (R) &= \frac{\pi R^2}{2} \Bigl( \mathcal{J}^T_{l_2m_{l_2},l_1m} (R) \mathcal{W}\left\{ \Phi_{n_1l_1m}^{A} , g_{l_1m} \right\}_R \\
& \quad \quad - \mathcal{N}^T_{l_2m_{l_2},l_1m} (R)
\mathcal{W}\left\{ \Phi_{n_1l_1m}^{A} , f_{l_1m} \right\}_R \Bigr),
\end{aligned}
\label{Eq_LFT}
\end{equation}
\textcolor{black}{and}
\begin{equation}
\begin{aligned}
\left\{
\begin{aligned}
\mathcal{J}^T_{l_2m_{l_2},l_1m} (R) &=
\sqrt{\frac{2l_2+1}{2l_1+1}}
\sum_{\beta=0}^\infty
i^{l_2+\beta -l_1} (2\beta +1)\,
C^{l_10}_{l_20,\beta 0}\,
C^{l_1m}_{l_2m_{l_2},\beta 0}\,
j_\beta (kR) \\[0.4em]
\mathcal{N}^T_{l_2m_{l_2},l_1m} (R) &=
\sqrt{\frac{2l_2+1}{2l_1+1}}
\sum_{\beta=0}^\infty
i^{l_2+\beta -l_1} (2\beta +1)\,
C^{l_10}_{l_20,\beta 0}\,
C^{l_1m}_{l_2m_{l_2},\beta 0}\,
y_\beta (kR).
\end{aligned}
\right.
\end{aligned}
\end{equation}
$\mathcal{W}\left\{\_ ,\_ \right\}_R$ symbolises the Wronskian evaluated at the internuclear distance $R$. $C^{JM}_{j_1m_1,j_2m_2}$ designates a Clebsch-Gordan coefficient and $j_\beta (kR)$ and $y_\beta (kR)$ are the usual regular and irregular spherical Bessel functions \cite{zotero-731}.

At this stage, both wavefunctions $\Phi_{n_1l_1m}^{A}$ and $\psi^{B^-}_{l_2m}$ are expressed relative to the same center. In addition, integrating Eq.~\eqref{eqref:interaction_term_plain} twice by parts, the volume integral can be rewritten as~\cite{robicheaux15,janev72}
\begin{equation}
\begin{aligned}
\bra{\Phi_{n_1l_1m}^{A}} V_{B} \ket{\psi^{B^-}_{l_2m}} &= \frac{\Xi^2}{2} \iint_\sigma d\Omega_\Xi \left[ \Phi_{n_1l_1m}^{\star A} (\textcolor{black}{\bm{\Xi}+\bm{R}}) \frac{\partial \psi^{B^-}_{l_2m}(\textcolor{black}{\bm{\xi}})}{\partial \xi} \Big|_{\bm{\Xi}}
- \psi^{B^-}_{l_2m} (\bm{\Xi}) \frac{\partial \Phi_{n_1l_1m}^{\star A}(\textcolor{black}{\bm{\Xi}+\bm{R}})}{\partial \xi}\Big|_{\bm{\Xi}}\right] \\
& \quad + \iiint_\tau d^3(\bm{\xi}) ~\psi^{B^-}_{l_2m} (\bm{\xi}) \left[ E^{B^-}\textcolor{black}{+\frac{\bm{\nabla_\xi}^2}{2}}\right] \Phi_{n_1l_1m}^{\star A}(\textcolor{black}{\bm{\xi}+\bm{R}}) .
\label{By_parts}
\end{aligned}
\end{equation}
The integration surface $\sigma$ is chosen to be a sphere centered on the perturber $B$, with a radius $\xi\equiv\Xi$ whose choice is discussed below. $\tau$ is the volume inside the surface $\sigma$.

\begin{figure}[h]
	\includegraphics[width=\columnwidth]{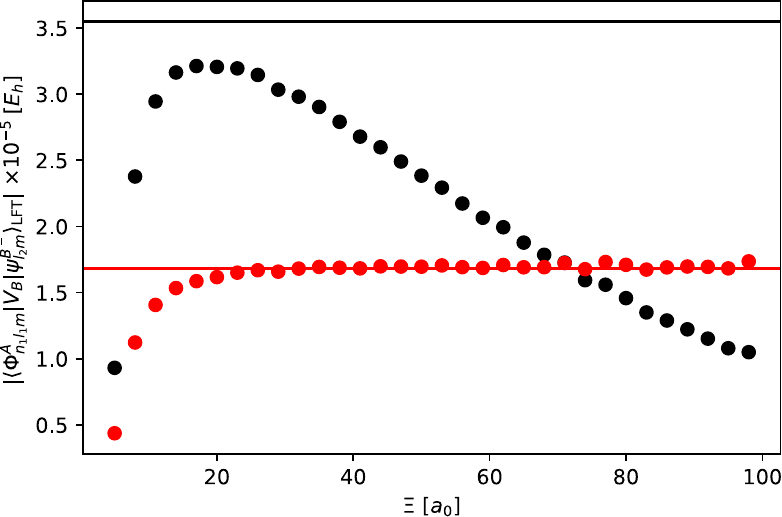}
	\caption{$\Xi$-evolution of the interaction calculated with the LFT method for the $12s$ Rydberg state, with $R=R_c$ (red full circles) and $R=190$~$a_0$ (black full circles). The horizontal lines correspond to the exact numerical calculation of the volume integral in both respective cases. \label{fig:LFT_Rc_vs_R}}
\end{figure}

The second term on the right hand side of Eq.~\eqref{By_parts} can be re-written, with the aid of~\eqref{eq:schrodinger_A}, as
\begin{equation}
\begin{aligned}
	\iiint_\tau d^3(\bm{\xi}) & ~\psi^{B^-}_{l_2m} (\bm{\xi})  \left[ E^{B^-}- E^A_{n_1l_1m} - 1/r\right]  \Phi_{n_1l_1m}^{\star A}(\textcolor{black}{\bm{\xi}+\bm{R}}) \simeq \\
	&\left[ E^{B^-}- E^A_{n_1l_1m} - 1/R\right]\iiint_\tau d^3(\bm{\xi})\psi^{B^-}_{l_2m} (\bm{\xi}) \Phi_{n_1l_1m}^{\star A}(\textcolor{black}{\bm{\xi}+\bm{R}}) ,
	\label{eq:volume_integral_simplified}
\end{aligned}
\end{equation}
where we chose the integration volume $\tau$ such that it contains entirely the region of space where the short-range potential $V_B$ is non negligible, while also being small enough such that the Coulomb potential of A$^+$ is approximately constant throughout. When $E^{B^-} - 1/R = E^A$, i.e., when the electron scatters resonantly off the atom B, the term vanishes. In the present case this occurs at the crossing internuclear distance $R_c$, at which point, combining Eqs. (\ref{Eq_LFT}) and (\ref{By_parts}), the interaction reduces to \textcolor{black}{the usual expression \citep{dochainarnaud,janev72,robicheaux15}}
\begin{equation}
\begin{aligned}
\bra{\Phi_{n_1l_1m}^{A}} V_{B} \ket{\psi^{B^-}_{l_2m}}_{\text{LFT}} \simeq& \frac{\Xi^2}{2} \mathcal{V}^T_{l_2m,l_1m}(R_c) \\
&\times \left[ f^{\star}_{l_2m} (\Xi) \frac{\partial \chi^{B^-}_{l_2m}(\xi)}{\partial \xi}\Big|_{\Xi}
- \chi^{B^-}_{l_2m} (\Xi) \frac{\partial f^{\star}_{l_2m}(\xi)}{\partial \xi}\Big|_{\Xi}\right].
\label{Surface_integral}
\end{aligned}
\end{equation}
The integration of angular variables can be carried out analytically and is contained in the $\mathcal{V}^T_{l_2m,l_1m}$ factor. We used the LFT index to label the matrix elements calculated with the LFT method.
Importantly, the above result is independent of the choice of $\Xi$ if $\Xi$ is
sufficiently large. Indeed, the radial anion wavefunction
$\chi^{B^-}_{l_2m}(\xi)$ asymptotically behaves as a spherical Bessel
function. The functions $\Phi_{n_1l_1m}^{\star A}(\bm{\xi})$ are also
expressed as linear combinations of spherical Bessel functions. The Wronskian of spherical Bessel functions with identical arguments, \textit{a fortiori} identical energies, is
proportional to $1/\Xi^2$, which implies
that Eq.~\eqref{Surface_integral} is independent of $\Xi$ (see red full circles in Fig. \ref{fig:LFT_Rc_vs_R}).  This is a crucial
point since it ensures that the calculated interaction term is independent of
how one chooses to divide space.

The above result however implies, reversibly, that,
for all internuclear distances but $R_c$, Eq.~\eqref{Surface_integral} depends
on the choice of $\Xi$. This is illustrated in Fig.~\ref{fig:LFT_Rc_vs_R} where the values for $R\neq R_c$ (full black circles) decrease monotonously at large $\Xi$. The reason is that, while the total matrix element should be constant, the integral~\eqref{eq:volume_integral_simplified} no longer vanishes
because the energy of the anionic state is different from the one of the ULRRM
state, i.e., the electron-scattering process we wish to describe is no longer
elastic. To include the $R$-dependence of the coupling in the description of
the electron transfer, Eq.~\eqref{Surface_integral} can no longer be used.

To determine the $R$-dependence of the interaction term, we have computed the volume integral contained in~\eqref{eqref:interaction_term_plain} directly, by transforming the coordinates of the wavefunction $\Phi^A_{n_1 l_1 m}$ using $\bm{r}= \bm{R} + \bm{\xi}$,
\begin{equation}
\bra{\Phi_{n_1l_1m}^{A}} V_{B} \ket{\psi^{B^-}_{l_2m}}_V= \iiint d^3 (\bm{\xi})~ \Phi_{n_1l_1m}^{\star A} (\bm{R}+\bm{\xi}) V_{B} (\xi)\psi^{B^-}_{l_2m} (\bm{\xi}).
\end{equation}
The index $V$ refers to matrix elements calculated with the direct numerical integration method.
Simpson integration was chosen \cite{virtanen20} and the radius of integration and number of grid points were systematically increased until reaching convergence. The integral converged within a radius $||\bm{\xi}||$=200~$a_0$ and 400, 100 and 100 grid points along the $\xi$ radial coordinate and the two angular coordinates, respectively, resulting in 4 million integration points in total. While the direct-integration approach is much more computationally demanding, with each computation lasting $>$ 10 minutes, it avoids the pitfalls of the approximations discussed above. The overlap integral was also computed numerically, with the same method and grid parameters.

\section{Results and discussion}
\label{Sec:results}
The interaction terms $\bra{\Phi_{n_1l_1m}^{A}} V_{B}
\ket{\psi^{B^-}_{l_2m}}$ calculated with the two methods outlined above
at $R=R_c$ are compared in Table~\ref{tab:interaction_vol_LFT}. At this
internuclear distance both methods are valid and the comparison can serve as a
benchmark of the accuracy of the two numerical calculations. The excellent
agreement between the results for different values of $n_1$ and $l_1$ makes us
confident that the calculations are well converged. The small discrepancy between the two methods is attributed to the limited numerical precision of the calculations and the small mismatch between the calculated and experimental electron affinities of Ca.

\sisetup{
    table-number-alignment = center
}

\begin{table}[h!]
\centering
\caption{Interaction (\ref{eqref:interaction_term_plain}) at the crossing internuclear distance $R_c$. The values are given in \textcolor{black}{$E_h$} and GHz, for both the numerical integration on the volume and the LFT method.}
\label{tab:interaction_vol_LFT}
\begin{tabular}{
l
l
S[table-format=-1.5]
S[table-format=-3.3]
S[table-format=-1.5]
S[table-format=-3.3]
}
\toprule
\multicolumn{2}{c}{\textbf{State}} &
\multicolumn{2}{c}{\textbf{$\bra{\Phi_{n_1l_1m}^{A}} V_{B} \ket{\psi^{B^-}_{l_2m}}_V$}} &
\multicolumn{2}{c}{\textbf{$\bra{\Phi_{n_1l_1m}^{A}} V_{B} \ket{\psi^{B^-}_{l_2m}}_{\text{LFT}}$}} \\
\cmidrule(lr){1-2}
\cmidrule(lr){3-4}
\cmidrule(lr){5-6}
\textbf{$n_1l_1$}
& \textbf{$n_1^\star$}
& {[$\times 10^{-5}$ \textcolor{black}{$E_h$}]}
& {[GHz]}
& {[$\times 10^{-5}$ \textcolor{black}{$E_h$}]}
& {[GHz]} \\
\midrule
12s & 9.66  & -1.68   & -111   & -1.68   & -111   \\
12f & 11.91 & -0.450  & -29.6  & -0.446  & -29.3  \\
14s & 11.66 & -0.275  & -18.1  & -0.279  & -18.4  \\
15f & 14.91 & -0.00367 & -0.241 & -0.00364 & -0.240 \\
\bottomrule
\end{tabular}
\end{table}

\begin{figure}[h]
	\includegraphics[width=\columnwidth]{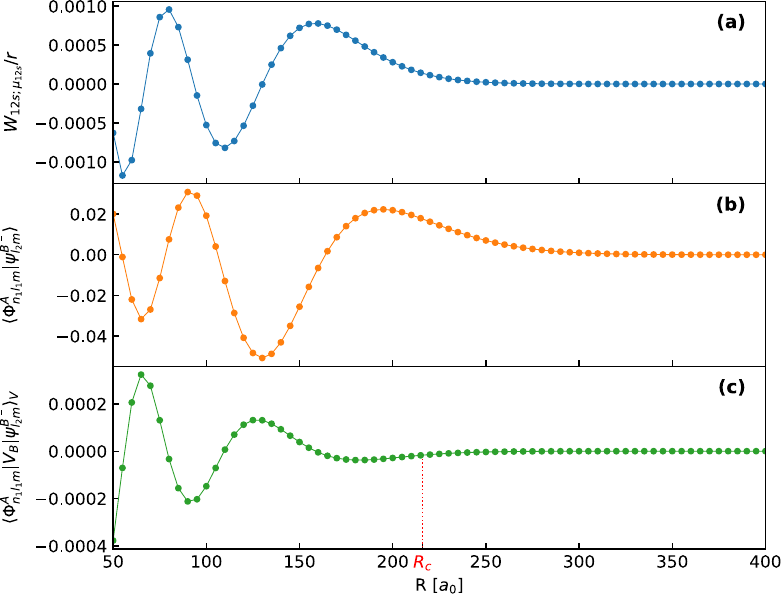}
	\caption{Comparison between the evolution of the interaction, the overlap and the Whittaker wavefunction with the internuclear distance $R$ for the $12s$ ULRRM state. All quantities are in a.u.. For the (a) panel, the $R$ variable of the x-axis is to be understood as $r$ as it is an atomic wavefunction. \label{fig:triple_plot}}
\end{figure}

The interaction strength calculated at $R \neq R_c$ with the direct method is
shown in Fig.~\ref{fig:triple_plot}(c) for $n_1=12$, $l_1=0$ and $m=0$. Its oscillations loosely follow the ones
of the Rydberg-electron wavefunction [Fig.~\ref{fig:triple_plot}(a)] and
increase in amplitude as $R$ becomes smaller, a fact that can be partly
attributed to an increased geometrical overlap of the wavefunctions as B gets closer to A$^+$.
In the region of the crossing ($R=R_c$), where the impact of electron transfer on molecular dynamics is the
largest, the interaction varies approximately linearly with $R$. Since the
slope of the interaction ($6.5\times 10^{-7}$~Hartree/$a_0$) is smaller than the
one of the energy difference between the two crossing potential-energy functions ($1/R_c^2 = 2.1 \times
10^{-5}$ Hartree/$a_0$), we expect that the interaction can be assumed to be
constant in the region of the crossing ($R \sim R_c$), an approximation often
made in electron-transfer calculations \cite{fabrikant00,fabrikant93,desfrancois95,dochain23,dochainarnaud}. In that case, the
surface-integral approach of Eq.~\eqref{Surface_integral} is well suited as its
low computational cost makes it possible to compute the ET interaction for a
broad range of $n_1$, $l_1$ and $m$ values.

The overlap integral shown in Fig.~\ref{fig:triple_plot}(b) reaches $s=0.018$ at
$R=R_c$. As outlined in Sec.~\ref{sec:problem_outline} this entails a
modification of the effective off-diagonal matrix element coupling
$\ket{\Phi^A_{n_1 l_1 m}}$ and $\ket{\psi^{B^-}_{n_2 l_2 m}}$ of the order of $s
(R-R_c) / R_c^2$ near the crossing. For $R-R_c = 10$~$a_0$, the change reaches
$5\times 10^{-6}$~Hartree, a quantity small compared to the value of the ET
interaction itself \textcolor{black}{($-1.68\times 10^{-5}$ $E_h$)}. The fact that the functions $\ket{\Phi^A_{n_1 l_1
m}}$ and $\ket{\psi^{B^-}_{n_2 l_2 m}}$ are not orthogonal can
therefore be neglected to within a good approximation.

The ET interaction is large for moderate values of $n_1$, reaching for example
$-1.68\times 10^{-5}$ Hartree ($111$~GHz, $457$~$\mu$eV, $3.69$~cm$^{-1}$) for
the $12s$ state at $R=R_c$, but falls rapidly as $n_1^*$ increases
(Table~\ref{tab:interaction_vol_LFT}). The crossing distance $R_c$ increases
with $n_1^*$ faster than the radius of the Rydberg-electron orbit (see Fig. \ref{fig:Energies}), such that
beyond $n_1 \gtrsim 15$ the wavefunctions of the Rydberg state ($\Phi^A$) and
anion state ($\psi^{B^-}$) no longer occupy the same regions of space and
electron exchange becomes unlikely. Experiments and theoretical calculations of
the electron transfer between Ne atoms excited to a Rydberg state and
ground-state calcium atoms show a similar behaviour and, in particular, a rapid decrease of the transfer rate for $n_1>12$~\cite{reicherts97,fabrikant00} that
becomes negligible beyond $n_1\sim 15$. Apart from an overall dephasing
associated to the quantum defect, the details of the short range potential
$V_{A^+}^\text{sh}$ have little influence on the long-range behaviour of the
Rydberg electron wavefunction and we expect the ET with Ne Rydberg atoms and Ca Rydberg atoms to be qualitatively similar.

The interactions considered so far were calculated assuming a single, isolated
Rydberg level described by the wavefunction $\Phi^A_{n_1 l_1 m}$, as in existing literature \cite{fabrikant00,fabrikant93,desfrancois95,dochain23,dochainarnaud}. However, the elastic scattering of the
Rydberg electron off the ground-state atom B, which leads to the formation of
the ULRRM potentials, mixes together $l_1$ and $n_1$ values. To take this into
account we computed the ET interaction between the ion pair and all ULRRM states $\ket{\psi^A_{m}}$ in the vicinity of $n_1^*=10$ and
with $m=0$. The results are listed in table~\ref{tab:interaction_Whit_ULRRM}
together with the interactions for isolated Rydberg levels with $l_1=0-9$. The
interactions involving ULRRM states are very similar to those for pure Rydberg
states, illustrating that (i) the interactions only weakly depend on $l_1$ such
that $l_1$-mixing has only a weak effect on the results; and (ii) the amount of
$l_1$-mixing, and to a greater extent $n_1$-mixing, induced by the Fermi
contact interaction for the states under consideration is relatively small. As
a result, table~\ref{tab:interaction_Whit_ULRRM} shows that the ET interaction
is relatively insensitive to the detailed composition of the ULRRM state and behaves much as in the case of isolated Rydberg levels.

The insensitivity described above does not imply, however, that electron-transfer dynamics have little influence on the structure and dynamics of ULRRMs. The strength of the interaction in the $n_1^* \sim 10$ range is comparable to, and
sometimes larger than, the well depth of ULRRM. An example is visible in
Fig.~\ref{fig:ULRRM}, where the depth of the outermost well of the molecular
state associated to the $12s$ Rydberg level (\textcolor{black}{12}~GHz) is \textcolor{black}{smaller than} the interaction strength (111~GHz, table~\ref{tab:interaction_vol_LFT}). The value of $R_c$ is 216 $a_0$, such
that the ULRRM and ion-pair energies are degenerate on the right wall of the ULRRM well. We thus expect the electron-transfer interaction to play a major role in
this range of internuclear distances and to significantly alter the structure and dynamics of the ULRRM vibronic levels. In particular, we expect near complete mixing between electronic states in a range of $\sim$~1~$a_0$ around the crossing distance $R_c$. However, since the electronic interaction term $\bra{\psi_{m}^{A}} V_{B}
\ket{\psi^{B^-}_{l_2m}}$ is \textcolor{black}{much larger than} the vibrational-level splitting, we cannot expect the dynamics to follow the Born-Oppenheimer picture used in the present work. Their description requires a nonadiabatic framework whose development is ongoing and whose details will be presented elsewhere.

\begin{table}[h!]
\centering
\caption{Comparison between the interactions, given by the LFT method and for the $n_1^\star = 10$ Rydberg state, in the case where the electronic wavefunction centered on the cation $A^+$ is a Whittaker function (pure atomic Rydberg state) or a linear combination of Whittaker functions (ULRRM).}
\label{tab:interaction_Whit_ULRRM}
\sisetup{table-format=1.3}
\begin{tabular}{ c S[table-format=1.3] S[table-format=1.3] }
\toprule
$l_1$ & \multicolumn{1}{c}{$\ket{\Phi_{n_1l_1m}^{A}}$} & \multicolumn{1}{c}{$\ket{\psi^{A}_m} = \sum_{nl} c_{nl}\ket{\Phi_{nlm}^{A}}$}  \\
     & \multicolumn{1}{c}{[$\times 10^{-5}$ \textcolor{black}{$E_h$}]} & \multicolumn{1}{c}{[$\times 10^{-5}$ \textcolor{black}{$E_h$}]} \\
\midrule
0 & 1.68 & \textcolor{black}{1.69} \\
1 & 1.89 & \textcolor{black}{1.88} \\
2 & 1.84 & \textcolor{black}{1.83} \\
3 & 2.90 & \textcolor{black}{2.91}  \\
4 & 2.65 & 2.65 \\
5 & 2.27 & \textcolor{black}{2.67}  \\
6 & 1.79 & \textcolor{black}{1.41}  \\
7 & 1.24 & \textcolor{black}{1.07}  \\
8 & 0.708 & \textcolor{black}{0.631}  \\
9 & 0.291 & \textcolor{black}{0.263}  \\
\bottomrule
\end{tabular}
\end{table}

Finally, we compared the values of the interaction obtained with the approximate anion wavefunction calculated in Sec. \ref{sec:anion_wvfn} ($-1.68\times 10^{-5} $~\textcolor{black}{$E_h$} for the 12$s$ state with $m=0$) with the ones from calculations using the anion wavefunction determined in \cite{dzuba97} by many-body perturbation theory for the $^2P_{1/2}$ state ($-1.06\times 10^{-5}$~\textcolor{black}{$E_h$} for the same state). The discrepancy of $\sim 40\%$ illustrates the degree of sensitivity of the interaction to the exact shape of the wavefunction. It gives a good estimate of the accuracy of our calculations, at present mainly limited by the accuracy of the description of the anion.

\section{Conclusion}

An electron can be resonantly exchanged between atoms excited to Rydberg states
with principal quantum numbers $n^*= 10 - 15$ and calcium ground-state atoms,
leading to the formation of ion pairs where Ca$^-$ is the negative ion. This
process is likely for relatively high Rydberg states because of the very low
electron affinity of Ca, making electron transfer efficient even at large internuclear distances. To describe the structure and dynamics of the transfer in the energy- and
internuclear-distance ranges where it is resonant or quasi-resonant, we have
calculated the electronic interaction between the neutral, Rydberg atom +
ground-state atom system, and the Ca$^+$ - Ca$^-$ ion pair. To
reach a quantitative description of the potential-energy landscape in the crossing region ($R \sim R_c$), the description of the former system includes
the effect of elastic zero-range scattering of the Rydberg electron on the
neutral target, leading to the formation of ultra-long-range Rydberg molecules.
Our calculations go beyond often-used approximations in order to determine the $R$-dependence of the interaction, falling in good agreement with those same approximate methods at $R=R_c$. For $n^* \sim 10$ the interaction is large, in
the 100~GHz range, which is sufficient to significantly disturb the potential-energy wells of the ULRRM. We show that the interaction depends only weakly on the approximate orbital angular-momentum quantum number, and is not significantly modified between isolated Rydberg levels and ULRRM states. It decreases rapidly with $n^*$ as a result of the increasing separation of the Rydberg- and negative-ion wavefunctions, such that it lies below the 100~MHz range for $n>15$.

The present results suggest that ULRRM states are significantly mixed with ion pairs in the $n^*=10-15$ range, where electron-transfer interactions are
expected to affect both the potential-energy landscape and vibrational
dynamics. The description of the nonadiabatic nuclear dynamics is
ongoing in our group. The present results represent an exciting perspective for future work as
they open a route toward the photoexcitation of ion pairs at ultracold
temperatures through their electronic mixing with ULRRM states, which have been
extensively produced in such conditions. While several theoretical proposals
exist for producing ion pairs in ultracold gases \cite{kirrander13,markson16,hummel20,peper20}, it has not been achieved
experimentally. The results given above consider two Ca atoms, but they can be generalized to other Rydberg systems straightforwardly, as only the
quantum defects included in the Whittaker functions and the electron-ground-state atom scattering phase shifts must be modified. The use of another neutral ground-state target with a low electron affinity, such as Sr (EA=52~meV) and Dy (EA=15~meV) species \cite{ning22}, is another possibility and could be described equally well with the approach outlined above.

\section*{Acknowledgements}
We are delighted to dedicate this article to Prof. Frédéric Merkt, whose groundbreaking work on Rydberg molecules and ion-pair states was a major source of inspiration for this work. MAGE also expresses his gratefulness for the support and profound influence Frédéric Merkt has had in shaping him as a scientist.

The authors gratefully acknowledge numerous fruitful discussion with X. Urbain and A. Dochain.

This work was supported by the Fonds de la Recherche Scientifique—FNRS under MIS Grant No. F.4027.24 and IISN Grant No. 4.4504.10. Computational resources have been provided by the supercomputing facilities of the Université catholique de Louvain (CISM/UCL) and the Consortium des Équipements de Calcul Intensif en Fédération Wallonie Bruxelles (CÉCI) funded by the Fond de la Recherche Scientifique de Belgique (F.R.S.-FNRS) under convention 2.5020.11 and by the Walloon Region.

\section*{Conflict of interest}
No potential competing interest was reported by the author(s).

\section*{Data availability statement}

The data that support the findings of this study are available from the corresponding author, M.G., upon reasonable request.

\bibliographystyle{tfo}

\end{document}